\documentclass[%
 reprint,
 amsmath,amssymb,
 aps,
]{revtex4-2}

\usepackage{graphicx}
\usepackage{dcolumn}
\usepackage{bm}
\usepackage{caption}
\usepackage{xcolor}
\usepackage{hyperref}
\usepackage{wasysym}

\begin{document}

\title{Boundary to bound dictionary for generic Kerr orbits}

\author{Riccardo Gonzo}
\email{rgonzo@ed.ac.uk}
\affiliation{Higgs Centre for Theoretical Physics, School of Physics and Astronomy, \\ University of Edinburgh, EH9 3FD, UK}%

\author{Canxin Shi}
\email{shicanxin@itp.ac.cn}
\affiliation{CAS Key Laboratory of Theoretical Physics, Institute of Theoretical Physics, \\ Chinese Academy of Sciences, Beijing 100190, China}%

\date{\today}

\begin{abstract}
We establish a new relation between classical observables for scattering and bound orbits of a massive probe particle in a Kerr background. We find an exact representation of the Hamilton-Jacobi action in terms of the conserved charges which admits an analytic continuation, both for the radial and polar contribution, for a general class of geodesics beyond the equatorial case. Remarkably, this allows to extend the boundary to bound dictionary and it provides an efficient method to compute the deflection angles and the time delay for scattering orbits, as well as frequency ratios for bound orbits, in the probe limit but at all orders in the perturbative expansion.
\end{abstract}

\maketitle

\section{Introduction}
\label{sec:intro}

The existence of gravitational waves was predicted by Einstein's theory of general relativity (GR) in 1916, but it took until 2015 for the Laser Interferometer Gravitational-Wave Observatory (LIGO) to detect the first direct evidence of these elusive waves \cite{LIGOScientific:2016aoc}. Since then, LIGO and other gravitational wave observatories around the world have detected numerous events, opening a new way to study the universe and test fundamental physics.

To accurately predict the properties of gravitational waves, a theoretical framework is required. One such framework is the post-Minkowskian (PM) expansion, which is a perturbative expansion in powers of the Newton constant $G_N$. Recent developments in the field of scattering amplitudes \cite{Damour:2017zjx,Neill:2013wsa,Damour:2019lcq,Bjerrum-Bohr:2018xdl,Cheung:2018wkq,Chung:2018kqs,Bern:2019crd,Bern:2020buy,Kosower:2018adc,Cristofoli:2021jas,Maybee:2019jus,Kalin:2020mvi,Mogull:2020sak,Bjerrum-Bohr:2022blt,Vines:2017hyw,Guevara:2018wpp,Guevara:2019fsj,Kosower:2022yvp} have pushed our understanding of the PM expansion for the classical two-body problem for spinless \cite{Bern:2019nnu,Bern:2021yeh,Dlapa:2021vgp,Dlapa:2022lmu,DiVecchia:2020ymx,DiVecchia:2021bdo,Bjerrum-Bohr:2021din,Bjerrum-Bohr:2021wwt,Brandhuber:2021eyq,Bjerrum-Bohr:2021wwt,Bini:2020wpo,Bini:2020nsb,Bini:2020hmy} and spinning bodies  \cite{Vines:2018gqi,Chung:2019duq,Arkani-Hamed:2019ymq,Kosmopoulos:2021zoq,Vines:2018gqi,Bern:2022kto,Jakobsen:2022zsx,Jakobsen:2022fcj,FebresCordero:2022jts,Chiodaroli:2021eug,Cangemi:2022bew,Alessio:2022kwv,Alessio:2023kgf,Chung:2020rrz,Liu:2021zxr,Aoude:2022trd,Aoude:2022thd,Chen:2021kxt,Menezes:2022tcs,Bautista:2023szu} up to high order for the conservative dynamics. The probe limit scenario is particularly relevant \cite{Cheung:2020gbf,Cheung:2020gyp,Gonzo:2021drq,Adamo:2022rmp,Bastianelli:2021nbs,Bautista:2021wfy,Bautista:2022wjf,Kol:2021jjc,Mino:2003yg,Poisson:2011nh,Harte:2011ku}, since it provides a concrete example of an exact resummation which makes contact with the self-force expansion \cite{Gralla:2008fgm,Barack:2018yvs}. For most of the cases, the Hamiltonian extracted from amplitudes can be directly fed into the effective-one-body machinery~\cite{Damour:1988mr,Buonanno:1998gg,Damour:1999cr,Damour:2008yg,Damour:2001tu} in order to generate gravitational wave templates for bound systems \cite{Khalil:2022ylj,Antonelli:2019ytb,Buonanno:2022pgc}.

Since the classical dynamics is completely captured by differential equations, only the boundary conditions provide the physical distinction between scattering and bound orbits. Building on such intuition, recently K\"alin and Porto \cite{Kalin:2019rwq,Kalin:2019inp} found a way to analytically continue scattering observables like the deflection angle into bound observables like the periastron advance \footnote{See also Damour and Deruelle \cite{Damour1985} for an interesting earlier proposal.}. This ``boundary to bound'' dictionary has been developed for two-body systems of spinless and aligned-spin particles, whose dynamics remain on the equatorial plane at all times. Recently, this has been partially extended to radiative observables \cite{Bini:2020hmy,Cho:2021arx,Saketh:2021sri,Jakobsen:2022zsx}. In the conservative case, one of the key insights in establishing such correspondence is given by the Hamilton-Jacobi action \cite{Damour:1988mr,Damour:1999cr,Leacock:1983,FORD1959259}, which is related to solution of the Bethe-Salpeter equation for classical bound states via the ``amplitude-action'' relation \cite{Bern:2021dqo,FORD1959259,Kol:2021jjc,Damgaard:2021ipf,Bjerrum-Bohr:2021wwt,Adamo:2022ooq}.

\begin{figure}[h!]
\includegraphics[width=0.68\linewidth]{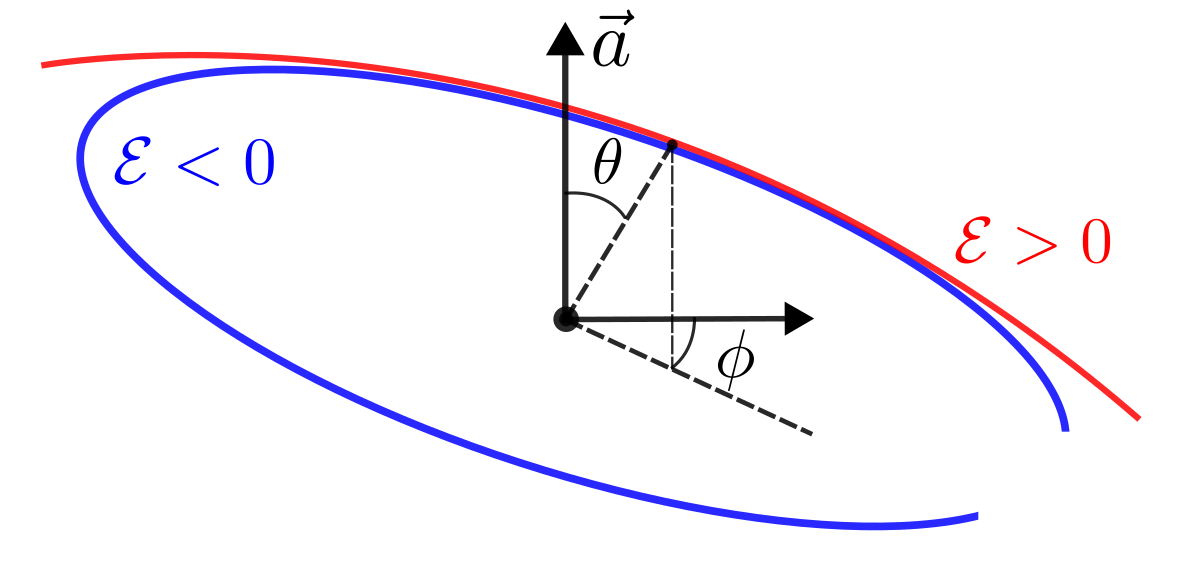}
\caption{We consider a class of orbits in a Kerr black hole which smoothly interpolates between scattering ($\mathcal{E}>0$, in red) and bound ($\mathcal{E}<0$, in blue) dynamics.}
\label{fig:Kerr-trajectory}
\end{figure}

Interestingly, the Hamilton-Jacobi action can be also used to describe massive probe particles moving in a Kerr metric beyond the equatorial case \cite{Carter:1968}. This raises the question of whether the boundary to bound dictionary can be extended to generic orbits. In this letter, we provide an affirmative answer to this question. We will show that there is a natural class of geodesics in a Kerr background which smoothly connect the scattering and the bound dynamics (see Fig.\ref{fig:Kerr-trajectory}), for which an analytic continuation is possible by taking into account also the Carter constant $Q$ beyond the energy $E$ and the projection of the angular momentum on the spin axis $L$. We will then derive the scattering angles $(\Delta\phi,\Delta\theta)$ and the time delay $\Delta T$ for time-like and null-like geodesics, respectively. Finally, using the new dictionary, we will compute the precession of the periastron $K^{\phi r}$ and of the orbital plane ${K^{\phi r}-K^{\theta r}}$, which are naturally expressed in terms of the fundamental frequencies $(\omega_r,\omega_{\phi},\omega_{\theta})$ of the motion \cite{Schmidt:2002qk,Fujita:2009bp}.

\paragraph*{Conventions} We use the mostly plus signature convention $(-+++)$ for the metric and we set $c=1$. {We also assume $L>0$ throughout the paper.}

\section{Hamilton-Jacobi action for generic Kerr orbits}
\label{sec:HJ-action}

The Kerr metric describes the spacetime of a spinning black hole of mass $M$ and spin $J= M a$. This can be written in Boyer-Lindquist coordinates $(t,r,\theta,\phi)$ as
\begin{align}
\label{eq:Kerr-BL}
\mathrm{d} s^2 &= - \frac{\Delta}{\Sigma} (\mathrm{d} t - a \sin^2(\theta) \mathrm{d} \phi)^2 + \frac{\Sigma}{\Delta} \mathrm{d} r^2 + \Sigma \, \mathrm{d} \theta^2 \\
&\quad+ \frac{\sin^2(\theta)}{\Sigma} [(r^2 + a^2) \mathrm{d} \phi - a \mathrm{d} t]^2 \,, \nonumber \\
\Delta(r) &= r^2 - 2 M r + a^2\,, \quad \Sigma(r,\theta) = r^2 + a^2 \cos^2(\theta) \,, \nonumber
\end{align}
where we have set Newton's constant to unity $G = 1$ and we have chosen the reference axis to be aligned with the spin direction.
The relativistic Hamiltonian for the geodesic motion of a probe particle of mass $m$ and 4-momentum $p_{\mu}$ in this metric is $H(x,p) = 1/2 \,g^{\mu \nu} p_{\mu} p_{\nu}$, which guarantees the validity of the geodesic equations
\begin{align}
\label{eq:geodesic}
p^{\mu} \nabla_{\mu} p^{\nu} = 0\,, \qquad g^{\mu \nu} p_{\mu} p_{\nu} = -m^2 \,.
\end{align}
A complete set of constants of motion can be determined for Kerr, as first shown by Carter \cite{Carter:1968}. First of all, the metric \eqref{eq:Kerr-BL} admits two Killing vectors $\partial_t$ and $\partial_{\phi}$ as a consequence of time-translation and axial symmetry. Therefore, the total energy $E$ and the angular momentum parallel to the spin axis $L$ as seen by an observer at spatial infinity are conserved
\begin{align}
\label{eq:conserved_EL}
E := - p_{\mu} \partial^{\mu}_t  = -p_t \,, \qquad
L := p_{\mu} \partial^{\mu}_{\phi} = p_{\phi} \,.
\end{align}
In addition to the isometries, the Kerr metric admits also an irreducible symmetric Killing tensor $K_{\mu \nu}$ which implies the existence of a new conserved charge  $Q$ called the Carter integral \footnote{The (positive definite) Carter constant is $k = K^{\mu \nu} p_{\mu} p_{\nu}$, but $Q$ is more convenient for our purposes.}
\begin{align}
\label{eq:conserved_Q}
Q &= K^{\mu \nu} p_{\mu} p_{\nu} - (L - a E)^2 \\
&= p_{\theta}^2 + a^2 (m^2 - p_t^2) \cos^2(\theta) + p_{\phi}^2 \cot^2(\theta) \,. \nonumber
\end{align}
Using a Euclidean flat space 3$d$ notation \cite{Balmelli:2015zsa,Khalil:2020mmr}, we can write $\vec{r} = r (\sin(\theta) \cos(\phi),\sin(\theta) \sin(\phi),\cos(\theta))$ and $\vec{a} = (0,0,a)$ so that we can suggestively recast \eqref{eq:conserved_Q} as
\begin{align}
Q &= |\vec{r} \wedge \vec{p}|^2 -(\vec{r} \wedge \vec{p} \cdot \hat{a})^2 - {(E^2-m^2)} (a \cdot \hat{r})^2 \\
&= |\vec{L}|^2 - L^2 - {(E^2-m^2)} (a \cdot \hat{r})^2  \,, \nonumber
\end{align}
i.e.{,} this is a measure of the motion of the particle off the equatorial plane given by the generalization of the equatorial projection of the orbital angular momentum $|\vec{L}|^2 - L^2$ for a spinning source \cite{Rosquist:2007uw}\footnote{For $a \to 0$, we recover $Q \stackrel{a \to 0}{\to} L_x^2 + L_y^2$ while $L \stackrel{a \to 0}{\to} L_z$. }.
For the scattering case, the relation between the conserved charges and the incoming kinematics is summarized in appendix~\ref{sec:AppendixD}.

The instantaneous 4-momentum $\mathcal{P} = p_{\mu} \mathrm{d} x^{\mu}$ of the probe particle can be now expressed \cite{Kapec:2019hro} in terms of the four constants of motion $(m^2,E,L,Q)$ by directly inverting the equations \eqref{eq:geodesic},\eqref{eq:conserved_EL} and \eqref{eq:conserved_Q}
\begin{align}
\label{eq:4-momentum}
\mathcal{P}(x) := - E \,\mathrm{d} t \pm_{r} \frac{\sqrt{R(r)}}{\Delta(r)} \,\mathrm{d} r \pm_{\theta} \sqrt{\Theta(\theta)} \,\mathrm{d} \theta + L \,\mathrm{d} \phi \,,
\end{align}
where we have defined the polar and radial potential
\begin{align}
\Theta(\theta) &:= Q + a^2 (E^2 - m^2) \cos^2(\theta) - L^2 \cot^2(\theta)\,, \\
R(r) &:= [E (r^2 + a^2) - a L]^2 - \Delta(Q + (L - a E)^2 + m^2 r^2)\,{;} \nonumber
\end{align}
the signs $\pm_{r}$ and $\pm_{\theta}$ depend on the radial and polar direction of the motion, respectively. For convenience, we choose $\pm_{r} = \pm_{\theta} = +$. The canonical 1-form \eqref{eq:4-momentum} provides the transformation to the principal function $S$, in terms of which the HJ action $I$ is defined as
\begin{align}
\label{eq:HJ_action}
&\qquad \quad I := \frac{S + E t - L \phi}{{2 \pi}} = I_r + I_{\theta} \,, \\
I_r &= \frac{1}{2 \pi} \int_{\mathcal{C}_r} p_r \, \mathrm{d} r \,, \qquad\,\,\,\,\,\,\,\,\,\,\, I_{\theta} = \frac{1}{2 \pi} \int_{\mathcal{C}_{\theta}} p_{\theta} \, \mathrm{d} \theta \,, \nonumber
\end{align}
where the paths $\mathcal{C}_r$ and $\mathcal{C}_{\theta}$ correspond to the physical trajectories for the radial and polar motion.
Since the dynamics in a Kerr spacetime is separable in Boyer-Lindquist coordinates, those contours $\mathcal{C}_k$ can be localized within the $(x^k, p_k)$ plane on the cotangent bundle.

\subsection{Boundary to bound dictionary for generic orbits}

We are interested in a class of generic orbits that smoothly connects the scattering and the bound regime. Generic geodesics are such that both endpoints are either a simple root of the radial potential $R(r)$, the horizon or infinity. The classification of time-like and null-like orbits in terms of the radial root structure has been recently completed in \cite{Compere:2021bkk} and \cite{Gralla:2019ceu}, respectively. We employ the conventions introduced in \cite{Compere:2021bkk} for the radial roots, which we review here. We use the symbols $\mid,+,-$ and $\rangle$ to label respectively the Kerr outer horizon, a region where motion is allowed $(R>0)$, a region where motion is disallowed $(R<0)$ and radial infinity, and the $\bullet$ to denote a single root. The radial root structure of the class of geodesics we are interested in is discussed in table \ref{tab:radialmotion}. \footnote{$E_{\text{ISCO}^+}$ stands for the energy of the prograde innermost stable circular orbit (ISCO). This will not be of further concern for our work, so we refer to \cite{Compere:2021bkk} for details.}
\begin{table}[h]
\begin{tabular}{|c|c|c|c|}
\hline Type & Energy range & Root structure & Radial range \\
\hline Unbound & $E > m$ & $|+ \bullet - {\color{red}\bullet} + \rangle$ & $r_m \leq r < \infty $ \\
\hline Bound & $E_{\text{ISCO}^+}< E < m$ & $|+ \bullet- {\color{red}\bullet} +{\color{blue}\bullet}-\rangle$ & $r_- \leq r \leq r_+ $ \\
\hline
\end{tabular}
\caption{The table shows the specific class of unbound orbits ($E > m$) with a single turning point $r_m$ (in red) that are smoothly connected with bound orbits ($E < m$) with two turning points $r_-$ (in red), $r_+$ (in blue).}
\label{tab:radialmotion}
\end{table}

At this point, we can define the cycle of integration for the Hamilton-Jacobi action for unbound and bound geodesics. We introduce the superscript $>$ to denote an expression valid for scattering orbits and $<$ to denote an expression valid for bound ones. For the radial motion, making manifest the dependence of the radial roots on the conserved charges, the radial integral becomes
\begin{align}
\int_{\mathcal{C}_r^{>}} = 2 \int_{r_m(\mathcal{E},l,a,l_Q)}^{\infty} \,, \qquad \int_{\mathcal{C}_r^{<}} = 2 \int_{r_-(\mathcal{E},l,a,l_Q)}^{r_+(\mathcal{E},l,a,l_Q)} \,,
\end{align}
where we have defined the conserved quantities per unit mass,
\begin{align}
\mathcal{E} := \frac{E^2 - m^2}{m^2} \,, \quad
l := \frac{L}{m} \,, \quad
l_Q := \frac{\sqrt{Q + L^2}}{m}{.}
\end{align}
A direct inspection of the analytic structure of the radial roots shows that we can generalize the boundary to bound dictionary from equatorial orbits \cite{Kalin:2019rwq,Kalin:2019inp} to generic orbits because of the remarkable map
\begin{align}
\label{eq:analy_radial}
r_-(\mathcal{E},l,a,l_Q) &\stackrel{\mathcal{E} < 0}{=}  r_m(\mathcal{E},l,a,l_Q)\,,  \\
r_+(\mathcal{E},l,a,l_Q) &\stackrel{\mathcal{E} < 0}{=}  r_m(\mathcal{E},-l,-a,-l_Q)\,, \nonumber
\end{align}
where $l_Q$ plays the role of an angular momentum.

For the polar motion, the condition $\Theta \geq 0$ implies that a generic geodesic with $l \neq 0$ is bounded between two turning points $\theta_- < \theta < \theta_+$, which are the solutions of the equation $\Theta = 0$ \footnote{We exclude the special case $l=0$ where the north ($\theta = {0}$) or the south ($\theta = {\pi}$) pole can be reached.}. The polar motion can be of ordinary or of vortical type according to the value of $\theta_\pm$, as shown in table \ref{tab:polarmotion}.

\begin{table}[h]
\begin{tabular}{|c|c|c|}
\hline Type & Polar range & Conditions \\
\hline Ordinary & $\theta_- < \pi/2 < \theta_+$, $\theta_- = \pi - \theta_+$ & $Q > 0, \mathcal{E} \lessgtr 0$\\
\hline Vortical & $\theta_- < \theta_+ < \pi/2$ or $\pi/2 < \theta_- < \theta_+$ & $Q \leq 0, \mathcal{E} > 0$\\
\hline
\end{tabular}
\caption{The table provides a qualitative classification of the polar motion.}
\label{tab:polarmotion}
\end{table}

Since we are interested in a class of geodesics that connect the unbound ($\mathcal{E} > 0$) and the bound ($\mathcal{E} < 0$) regime, we are forced to restrict to the case of the oscillatory polar motion with $Q > 0$.
After excluding the degenerate case of planar geodesics at fixed polar angle $\theta = \theta_{\pm}$, the angular integral for the generic configuration reads \cite{Kapec:2019hro}
\begin{align}
\int_{\mathcal{C}_{\theta}} = 2 n \Bigg|\int_{\pi/2}^{\theta_{\pm}}\Bigg| + \eta_{\text{in}} \Bigg|\int_{\pi/2}^{\theta_{\text{in}}}\Bigg| - \eta_{\text{out}} \Bigg|\int_{\pi/2}^{\theta_{\text{out}}}\Bigg|\,,
\label{eq:countourpolargeneric}
\end{align}
where $\theta_{\text{in}}$ (resp. $\theta_{\text{out}}$) is the initial (resp. final) polar angle of the trajectory, $n$ is the number of turning points of the polar motion and we have defined the signs
\begin{align}
\eta_{\text{in}/\text{out}} &= \text{sign}(p_{\text{in}/\text{out}}^{\theta}) \, \text{sign}(\cos(\theta_{\text{in}/\text{out}})) \,.
\end{align}
We now consider the class of geodesics which start on the equatorial plane with $\theta_{\text{in}} = \pi/2$, which is a convenient simplification of our problem and will not affect the validity of the analytic continuation. With this choice, the physical observables we will compute depend only on the conserved charges. Therefore, we can effectively use
\begin{align}
\label{eq:contourpolar}
\int_{\mathcal{C}_{\theta}} \to 2 n \Bigg|\int_{\pi/2}^{\theta_{\pm}}\Bigg|  - \eta_{\text{out}} \Bigg|\int_{\pi/2}^{\theta_{\text{out}}}\Bigg|\,,
\end{align}
where $\theta_{\text{out}}$ will be determined explicitly in terms of the conserved charges, as we will discuss later.

We are now ready to compute the Hamilton-Jacobi action for our class of geodesics. We start with the radial action, which we can write for scattering orbits as
\begin{align}
\label{eq:radialaction1}
I_r^{>} &:= \frac{1}{2 \pi} \int_{\mathcal{C}_r^{>}} p_r \, \mathrm{d} r \stackrel{u=1/r}{=} \frac{1}{\pi} \int_0^{u_m} \frac{\mathrm{d} u}{u^2} \frac{\sqrt{R(u)}}{\Delta(u)}  \\
&= \frac{1}{2 \pi} \frac{{m} \sqrt{\mathcal{E}}}{\sqrt{M^2 - a^2}} \nonumber\\
& \,\,\times \int_0^{u_m} \frac{\mathrm{d} u}{u^2} \prod_{j=1}^4 \left(1 - \frac{u}{u_j}\right)^{\frac{1}{2}} \left(\frac{1}{u_B - u} - \frac{1}{u_A - u}\right) \,, \nonumber
\end{align}
where we have defined the radial roots $\{u_j\}_{j=1,\dots,4}$ and
\begin{align}
R(u) = -&\frac{a^2 Q}{u^4} \prod_{j=1}^4 (u - u_j)\,, \\
u_{A} = \frac{M + \sqrt{M^2 - a^2}}{a^2}\,, &\,\,\, u_{B} = \frac{M - \sqrt{M^2 - a^2}}{a^2} \,. \nonumber
\end{align}
Having selected the radial root corresponding to the minimum distance according to the pattern identified in table~\ref{tab:radialmotion}, say $u_4 = u_m$, we can then change variables to $u = u_m \tilde{u}$ so that the radial action reads
\begin{align}
\label{eq:radialaction2}
I_r^{>,\epsilon} &= \frac{1}{2 \pi} \frac{m \sqrt{\mathcal{E}}}{u_m^{1-\epsilon} \sqrt{M^2 - a^2}} \int_0^{1} \frac{\mathrm{d} \tilde{u}}{\tilde{u}^{2-\epsilon}}   \\
& \times \prod_{j=1}^4 \left(1 - \frac{u_m}{u_j} \tilde{u }\right)^{\frac{1}{2}} \left[\frac{1}{u_B-u_m \tilde{u}} - \frac{1}{u_A-u_m \tilde{u}}\right] \,,\nonumber
\end{align}
where we have introduced an infrared regulator ${\operatorname{Re}\epsilon > 1}$ to make the integral well-defined \cite{Damour:1988mr,Damour:2019lcq}. We can then provide a closed-form expression for the radial action in terms of the Lauricella hypergeometric functions $F_D^{(n)}$,
\begin{align}
\label{eq:radialaction3}
&I_r^{>,\epsilon} = \frac{1}{2 \pi} \frac{m \sqrt{\mathcal{E}}}{u_m^{1-\epsilon} \sqrt{M^2 - a^2}} \frac{\Gamma(-1+\epsilon) \Gamma(3/2)}{\Gamma(1/2+\epsilon)}  \\
&\times\! \bigg[ \frac{1}{u_B} F_D^{(4)}\bigg(\alpha_r,\vec{\beta}_r,\gamma_r; \frac{u_m}{u_B},\frac{u_m}{u_1},\frac{u_m}{u_2},\frac{u_m}{u_3}\bigg)
\!-\! (u_B \!\leftrightarrow\! u_A) \bigg]\nonumber \\
&\alpha_r=-1+\epsilon, \quad \vec{\beta}_r = \Big\{1,-\frac{1}{2},-\frac{1}{2},-\frac{1}{2}\Big\}, \quad \gamma_r=\frac{1}{2}+ \epsilon \,. \nonumber
\end{align}
For bound orbits, we can use the relation \eqref{eq:analy_radial} to write the radial contour as
\begin{align}
\int_{\mathcal{C}_r^{<}} &\stackrel{\mathcal{E} < 0}{=} {2}\int_{r_-(\mathcal{E},l,a,l_Q)}^{\infty} - {2}\int_{r_+(\mathcal{E},l,a,l_Q)}^{\infty} \\
&\stackrel{\mathcal{E} < 0}{=} {2}\int_{r_m(\mathcal{E},l,a,l_Q)}^{\infty} - {2}\int_{r_m(\mathcal{E},-l,-a,-l_Q)}^{\infty}\,. \nonumber
\end{align}
Since $p_r$ is invariant under $(a,l,l_Q) \to (-a,-l,-l_Q)$, we can establish the analytic continuation
\begin{align}
\label{eq:Ir_B2B}
&I_r^{<,\epsilon}(\mathcal{E},l,a,l_Q)  \\
& \quad \stackrel{\mathcal{E}<0}{=} I_r^{>,\epsilon}(\mathcal{E},l,a,l_Q) - I_r^{>,\epsilon}(\mathcal{E},-l,-a,-l_Q) \,. \nonumber
\end{align}
At this point, we focus on the polar action and we compute separately the contribution of both terms in \eqref{eq:contourpolar}. {On the branches shown in Fig.\ref{fig:polar_angle}, once} we choose the initial condition $\text{sign}(p_{\text{in}}^{\theta}) = 1$, the equations of motion will impose (see appendix \ref{sec:AppendixA})
\begin{align}
\label{eq:prefactors_polar}
\eta^{>}_{\text{out}} = -1\,, \quad n^{>} = 1\,, \qquad \eta^{<}_{\text{out}} = {-1}\,, \quad n^{<} = 2\,,
\end{align}
for scattering and bound orbits, respectively. The first contribution for scattering orbits reads
\begin{align}
\label{eq:polaraction1}
I_{\theta}^{> (1)} &:= \frac{1}{2 \pi} \int_{\mathcal{C}_{\theta}^{>}} p_{\theta} \, \mathrm{d} \theta \Bigg|_1 = \frac{n^{>}}{\pi} \int_{\pi/2}^{\theta_{+}} \mathrm{d} \theta \, \sqrt{\Theta(\theta)} \\
&\!\!\!\!\!\!\! \stackrel{\cos^2(\theta)=U_{+} \tilde{U}}{=} \frac{\sqrt{Q U_{+}}}{2 \pi}
\int_0^{1} \frac{\mathrm{d} \tilde{U} (1-\tilde{U})^{\frac{1}{2}} }
{\tilde{U}^{\frac{1}{2}} (1 - U_+ \tilde{U})} \left(1-\frac{U_{+}}{U_-}\tilde{U} \right)^{\frac{1}{2}} \,, \nonumber
\end{align}
where the roots $U_{\pm}$ of the polar potential are \footnote{The definition of $U_{\pm}$ here is slightly different than the conventional one \cite{Kapec:2019hro}, and it allows for a smooth analytic continuation for $\mathcal{E}<0$.}
\begin{align}
U_{\pm} \!&=\! \frac{\Delta U \!\pm\! \sqrt{\Delta U^2 \!+\! a^2 \mathcal{E} (l_Q^2 - l^2)}}{a^2 \mathcal{E}}\,, \\
& \qquad \Delta U = \frac{ a^2 \mathcal{E} - l_Q^2}{2}\,, \nonumber
\end{align}
and it is possible to show that $\theta_{\mp} = \arccos(\pm \sqrt{U_{+}})$. Therefore, the first term can be written in terms of the Lauricella hypergeometric function $F_D^{(q)}$, \footnote{In the case for $q=2$, Lauricella function $F_D^{(q)}$ reduces to Appell's $F_1$ hypergeometric series.}
\begin{align}
\label{eq:polaraction2}
I_{\theta}^{> (1)} &= \frac{1}{4} \sqrt{Q U_{+}} F_D^{(2)}\left(\alpha^{(1)}_{\theta},\vec{\beta}^{(1)}_{\theta}, \gamma^{(1)}_{\theta} ; U_+, \frac{U_+}{U_-}\right) \,,  \\
&\alpha^{(1)}_{\theta}=\frac{1}{2}, \quad \vec{\beta}^{(1)}_{\theta} = \Big\{1,-\frac{1}{2}\Big\}, \quad \gamma^{(1)}_{\theta}=2 \,. \nonumber
\end{align}
Therefore{,} the second term in \eqref{eq:contourpolar} can be written as
\begin{align}
\label{eq:polaraction3}
&I_{\theta}^{> (2)} = \frac{1}{2 \pi} \int_{\mathcal{C}_{\theta}^{>}} p_{\theta} \, \mathrm{d} \theta \Bigg|_2 = -\frac{\eta_{\text{out}}^{>}}{2 \pi} {\Bigg|}\int_{\pi/2}^{\theta_{\text{out}}^{>}} \mathrm{d} \theta \, \sqrt{\Theta(\theta)}{\Bigg|} \\
&= \frac{1}{2 \pi} \sqrt{Q U^{>}_{\text{out}}} F_D^{(3)}\left(\alpha^{(2)}_{\theta},\vec{\beta}^{(2)}_{\theta}, \gamma^{(2)}_{\theta} ; U^{>}_{\text{out}}, \frac{U^{>}_{\text{out}}}{U_+}, \frac{U^{>}_{\text{out}}}{U_-}\right) \,,\nonumber \\
&\qquad \alpha^{(2)}_{\theta}=\frac{1}{2}, \quad \vec{\beta}^{(2)}_{\theta} = \Big\{1,-\frac{1}{2},-\frac{1}{2}\Big\}, \quad \gamma^{(2)}_{\theta}=\frac{3}{2} \,, \nonumber
\end{align}
where the $U^{>}_{\text{out}}$ is determined from (see \cite{Kapec:2019hro})
\begin{align}
\label{eq:Uout}
U^{>}_{\text{out}}&= U_{+} \operatorname{sn}^2\left(X_0^{>} \Big| {\frac{U_{+}}{U_{-}}}\right) \,, \\
X_0^{>}&=-\sqrt{- a^2 m^2 \mathcal{E} U_{-}} \int_{\mathcal{C}_r^{>}} \frac{\mathrm{d} r}{ \sqrt{{R}(r)}} \,. \nonumber
\end{align}
For bound orbits, the turning points of the polar potential are still $\theta_{\mp} = \arccos(\pm \sqrt{U_{+}})$, so we only need to perform the conventional analytical continuation of the radial contour $\mathcal{C}_r^{>} \to \mathcal{C}_r^{<}$ for $U_{\text{out}}$ in \eqref{eq:Uout}. The polar action for $\mathcal{E} < 0$ will therefore be
\begin{align}
\label{eq:Itheta_B2B}
I^{<}_{\theta} (\mathcal{E},l,a,l_Q;&n^{>},\eta_{\text{out}}^{>})
\stackrel{\mathcal{E}<0}{=} I^{>}_{\theta} (\mathcal{E},l,a,l_Q;n^{<},\eta_{\text{out}}^{<})  \,,
\end{align}
where we emphasize again that $n^{>} = 1 \to n^{<} = 2$ and $\eta^{>}_{\text{out}} = -1 \to \eta^{<}_{\text{out}} = {-1}$ {on the branches considered here.}

\section{Perturbative expansion of scattering observables}
\label{sec:scatteringPM}

In this section we derive the post-Minkowskian (PM) expansion of the scattering angles $(\Delta\theta,\Delta\phi)$ and the time delay $\Delta T$ by using the equations of motion coming from the HJ action \eqref{eq:HJ_action}. Since our main focus is on the weak field limit, we present our results as a double expansion in $M$ and $a$ with $a \ll M$.
We refer the reader to appendix \ref{sec:AppendixB} for the exact expression of scattering angles in terms of hypergeometric functions.
\vspace{-13pt}
\subsection{Polar deflection angle $\Delta\theta$}

The polar deflection angle $\Delta\theta = \theta_{\text{out}} - (\pi - \theta_{\text{in}})$ is completely determined by $U_{\text{out}}=\cos^2(\theta_{\text{out}})$, given that we set $\theta_{\text{in}} = \pi/2$. It is straightforward to extend all our calculations to a generic incoming angle $\theta_{\text{in}}$, for example by using the generic polar contour \eqref{eq:countourpolargeneric} in the $r-\theta$ equation \eqref{eq:rtheta_eq} of appendix \ref{sec:AppendixA}. A direct perturbative expansion of $\Delta\theta$ from \eqref{eq:Uout} gives, up to $\mathcal{O}(M^3 a^2)$,
\allowdisplaybreaks
\begin{align}
\label{eq:polar_angle}
&\frac{\Delta\theta}{\sqrt{l_Q^2 - l^2}}  =  -\frac{2 M (2 \mathcal{E}+1)}{\sqrt{\mathcal{E}} \, l_Q^2} -\frac{3 \pi M^2 (5 \mathcal{E}+4)}{4\, l_Q^3} \\
&\qquad -\frac{2 M^3 (64 \mathcal{E}^3+72 \mathcal{E}^2+12 \mathcal{E} \!-\!1)}{3 \,\mathcal{E}^{3/2}\, l_Q^4}
+\frac{4 M^3 l^2 (2 \mathcal{E}+1)^3}{3 \,\mathcal{E}^{3/2}\, l_Q^6}\nonumber \\
&\qquad+ a \sqrt{\mathcal{E}+1} \Bigg(\frac{8 M l \sqrt{\mathcal{E}}}{l_Q^4}
+ \frac{3 \pi M^2 l (5 \mathcal{E}+2)}{l_Q^5} \nonumber \\
&\qquad- \frac{16 M^3 l^3 (2 \mathcal{E}+1)^2}{\sqrt{\mathcal{E}}\, l_Q^8}
+\frac{16 M^3 l (16 \mathcal{E}^2+12 \mathcal{E}+1)}{\sqrt{\mathcal{E}} \,l_Q^6} \Bigg) \nonumber \\
&+ a^2 \Bigg(\frac{2 \sqrt{\mathcal{E}} M (l_Q^2-4\, l^2) (2 \mathcal{E}+1)}{l_Q^6} \nonumber \\
&\qquad+\frac{3 \pi M^2 (l_Q^2-5\, l^2) (95 \mathcal{E}^2+88 \mathcal{E}+8)}{32 \, l_Q^7} \nonumber \\
&\qquad+\frac{16 M^3 l^4 (2 \mathcal{E}+1) (8 \mathcal{E}^2+ 8 \mathcal{E}+1) }{\sqrt{\mathcal{E}}\, l_Q^{10}} \nonumber \\
&\qquad-\frac{12 M^3 l^2 (88 \mathcal{E}^3+116 \mathcal{E}^2+34 \mathcal{E}+1) }{\sqrt{\mathcal{E}}\, l_Q^8} \nonumber \\
&\qquad+\frac{8 M^3 \sqrt{\mathcal{E}} (20 \mathcal{E}^2+26 \mathcal{E}+7)}{l_Q^6}\Bigg) \,. \nonumber
\end{align}

\subsection{Azimuthal deflection angle $\Delta\phi$}

The azimuthal scattering angle $\Delta\phi$ is the conjugate variable to the angular momentum $L$, i.e.{,}
\begin{align}
\label{eq:phi_angle}
\frac{\Delta\phi + \pi}{2 \pi} = -\frac{\partial I}{\partial L} = -\frac{\partial I_r}{\partial L} -\frac{\partial I_{\theta}}{\partial L} \,,
\end{align}
which can be computed from the Hamilton-Jacobi action in \eqref{eq:radialaction3}, \eqref{eq:polaraction1} and \eqref{eq:polaraction2}. It is worth stressing that we need to keep $U^{>}_\text{out}$ invariant in taking the derivative over $L$, since technically it is only fixed dynamically by the equations of motion, whereas the HJ action works at the off-shell level~\footnote{One can also derive $\Delta \phi$ from Hamilton's principal function, which should be understood as a type-2 generating function for a canonical transformation.}. A direct calculation up to order $\mathcal{O}(M^3 a^2)$ in the PM expansion gives
\begin{align}
&\Delta\phi = \frac{2 M l (2 \mathcal{E}+1)}{\sqrt{\mathcal{E}} \, l_Q^2}+\frac{3 \pi M^2 l (5 \mathcal{E}+4)}{4 \, l_Q^3}  \\
&\qquad +\frac{2 M^3 l (64 \mathcal{E}^3+72 \mathcal{E}^2+12 \mathcal{E}-1)}{3 \, \mathcal{E}^{3/2} \, l_Q^4} \nonumber \\
&+ a \sqrt{\mathcal{E}+1} \Bigg(\frac{4 M \sqrt{\mathcal{E}} (l_Q^2-2 l^2)}{l_Q^4} \nonumber \\
&\qquad +\frac{\pi M^2 (5 \mathcal{E}+2) (l_Q^2-3 l^2)}{l_Q^5}\nonumber \\
&\qquad + \frac{4 M^3 (16 \mathcal{E}^2+12 \mathcal{E}+1) (l_Q^2-4 l^2)}{\sqrt{\mathcal{E}}\, l_Q^6} \Bigg) \nonumber \\
&+a^2 \Bigg(\frac{2 M {l} \sqrt{\mathcal{E}} (2 \mathcal{E}+1) (4 l^2-3 l_Q^2)}{l_Q^6} \nonumber \\
& \qquad +\frac{3 \pi M^2 l (95 \mathcal{E}^2+88 \mathcal{E}+8) (5 l^2-3 l_Q^2)}{32 \, l_Q^7} \nonumber \\
& \qquad+\frac{4 M^3 l (128 \mathcal{E}^3+168 \mathcal{E}^2+48 \mathcal{E}+1) (2 l^2-l_Q^2)}{\sqrt{\mathcal{E}}\, l_Q^8} \Bigg) \nonumber \\
& + M^3 (l_Q^2 - l^2) \Bigg[\frac{8 l (2 \mathcal{E}+1)^3}{3 \, \mathcal{E}^{3/2} \, l_Q^6} -\frac{32 a\, l^2 \sqrt{\mathcal{E}+1} (2 \mathcal{E}+1)^2}{\sqrt{\mathcal{E}}\, l_Q^8} \nonumber \\
& \quad +\frac{8 a^2 l (2 \mathcal{E}+1) \left(4  l^2 \left(8 \mathcal{E}^2+8 \mathcal{E}+1\right)-(2 \mathcal{E}+1)^2 l_Q^2\right)}{\sqrt{\mathcal{E}}\, l_Q^{10}}\Bigg]\,,\nonumber
\end{align}
where we have isolated with the last square bracket the contributions from the polar action, which are proportional to $Q = m^2 (l_Q^2 -l^2)$. It is possible to notice a simple relation between the angles $\Delta\phi$ and $\Delta\theta$ as $a \to 0$, i.e.{,} $\Delta\phi \stackrel{a \to 0}{\sim} -(l/\sqrt{l_Q^2 -l^2}) \Delta\theta$ at the lowest order, which is determined by the fact that the motion happens on an inclined plane and there is always a change of coordinates to bring it to the standard equatorial plane. Moreover, in the limit $l_Q \to l$, we recover as expected the well-known equatorial expression \cite{Vines:2018gqi,Damgaard:2022jem}.

\subsection{Time delay}

The time delay is related to the conjugate variable of the energy $E$ in the HJ action, but it is defined only when we compare the measure to an observer at large distances \cite{Camanho:2014apa,AccettulliHuber:2020oou,Bautista:2021wfy,Bellazzini:2022wzv}.
Having defined the impact parameter $b$ for generic null geodesics
\begin{align}
r_m \stackrel{M \to 0}{=} b = |\vec{b}| \longrightarrow b = \frac{\sqrt{L^2 + Q}}{E} \,,
\end{align}
and the effective inclination angle \cite{Ryan:1995zm}
\begin{align}
\cos(\iota) = \frac{L}{\sqrt{L^2 + Q}} \,,
\end{align}
we can then compute the time delay $\Delta T$ for generic null geodesics with fixed $b$ relative to an observer with $b' \gg b$ but at the same energy $E' = E$,
\begin{align}
\Delta T &= {2\pi}\left[\frac{\partial I}{\partial E} \Bigg|_{b,E} - \frac{\partial I}{\partial E} \Bigg|_{b' \gg b,E'=E}\right]  \\
&= 4 M \log\left(\frac{b'}{b}\right) +\frac{15 \pi M^2}{2 \, b} + \frac{64 M^3}{b^2} \nonumber \\
&-\frac{a \cos(\iota)}{b} \left(8 M + \frac{15 \pi M^2}{b}+\frac{256 M^3}{b^2}\right) \nonumber \\
&+ \frac{a^2}{b^2} \Big(6 M \cos(2 \iota)+ \frac{95 \pi M^2}{16\, b} (1+3 \cos(2 \iota))\nonumber \\
& \qquad\qquad\qquad+\frac{32 M^3}{b^2} (7+13 \cos(2 \iota))\Big)\,, \nonumber
\end{align}
which is accurate up to order $\mathcal{O}(M^3 a^2)$. {Here the energy derivatives are taken at fixed $L$ and $Q$.} As expected, $\Delta T$ is positive because of causality arguments \cite{Camanho:2014apa} as long as we impose the physical condition $|a| \leq M$.

\section{Perturbative expansion of bound observables}
\label{sec:boundPM}

Using the boundary to bound dictionary for the Hamilton-Jacobi action developed in \eqref{eq:Ir_B2B} and \eqref{eq:Itheta_B2B}, we now proceed to compute the perturbative expansion of bound observables for generic bound orbits which are connected to scattering ones via analytic continuation. We use the same conventions as in section \ref{sec:scatteringPM}.

\subsection{The fundamental frequencies $\omega_r, \omega_{\phi}, \omega_{\theta}$}

The basic properties of Kerr bound orbits are specified by the so-called fundamental frequencies~\cite{Schmidt:2002qk, Hinderer:2008dm, Kerachian:2023oiw}.
Although they are coordinate independent, it is useful to describe them via the conjugate momenta of the action-angle variables in the Boyer-Lindquist representation,
\begin{gather}
    J_t := E, \qquad
    J_r := \frac{1}{2 \pi} \oint p_r \, \mathrm{d} r = I_r^{<},  \\
    J_\theta := \frac{1}{2 \pi} \oint p_\theta \mathrm{d} \theta = I_{\theta}^{< (1)}, \quad
    J_\phi := \frac{1}{2 \pi} \oint p_\phi \mathrm{d} \phi = L. \nonumber
\end{gather}
Since the coordinates are integrated out, the action momenta are constants of the motion $J_\beta (H, E, L, Q)$, where $H = -m^2/2$ is the Hamiltonian for the action-angle variables.
The fundamental frequencies are defined as
\begin{align}
    \omega_r = \frac{\partial H}{\partial J_r}\,, \quad
    \omega_{\theta} = \frac{\partial H}{\partial J_{\theta}}\,, \quad
    \omega_{\phi} = \frac{\partial H}{\partial J_{\phi}}\,.
\end{align}
Note that the partial derivative are taken with $J_\beta$ being invariant.
Using the results in appendix \ref{sec:AppendixC}, we can express these frequencies as
\begin{gather}
    \label{eq:fund_frequencies}
    \omega_r = -\frac{1}{\Omega} \frac{\partial J_{\theta}}{\partial Q}\,, \quad
    \omega_{\theta} = \frac{1}{\Omega} \frac{\partial J_{r}}{\partial Q} \,,
     \\
    \omega_{\phi} = \frac{1}{\Omega} \bigg( \frac{\partial J_{r}}{\partial L} \frac{\partial J_{\theta}}{\partial Q} - \frac{\partial J_{r}}{\partial Q} \frac{\partial J_{\theta}}{\partial L} \bigg) \,, \nonumber
\end{gather}
with $\Omega := {\frac{\partial J_{r}}{\partial Q} \frac{\partial J_{\theta}}{\partial H}-\frac{\partial J_{r}}{\partial H} \frac{\partial J_{\theta}}{\partial Q}}$.
These frequencies have been computed in a closed form in \cite{Schmidt:2002qk,Fujita:2009bp}{. Their values depend on the time parametrization.} Therefore{,} the natural bound observables are the frequency ratios \cite{Lewis:2016lgx}
\begin{gather}
\label{eq:ratiophi}
K^{\phi r} := \frac{\omega_{\phi}}{\omega_r} = \frac{\partial J_{r}/\partial Q}{\partial J_{\theta}/\partial Q}\, \frac{\partial J_{\theta}}{\partial L} -\frac{\partial J_{r}}{\partial L}, \\
\label{eq:ratiotheta}
K^{\theta r} := \frac{\omega_{\theta}}{\omega_r} = -\frac{\partial J_{r}/\partial Q}{\partial J_{\theta}/\partial Q}\,,
\end{gather}
which are related to the precession rate of the periastron ($K^{\phi r}$) and of the orbital plane (${K^{\phi r}-K^{\theta r}}$).

\subsection{The periastron precession rate $\omega_{\phi}/\omega_r$}

In the weak field regime, we can identify the precession rate of the orbital ellipse with the so-called periastron advance rate $K^{\phi r}$.
A direct calculation of \eqref{eq:ratiophi} gives, up to order $\mathcal{O}(M^3 a^2)$ in the weak-field expansion,
\begin{align}
\label{eq:Kphir}
K^{\phi r} &=1 + \frac{3 M^2 (5 \mathcal{E}+4)}{4\, l_Q^2} \\
&\quad + \frac{a\, M^2 \sqrt{\mathcal{E}+1} (l_Q-3 l) (5 \mathcal{E}+2)}{l_Q^4} \nonumber \\
&\quad + \frac{3 a^2 M^2}{32 \, l_Q^6} \Big[ l^2 (445 \mathcal{E}^2+416 \mathcal{E}+40)  \nonumber \\
& \qquad  \qquad \qquad - l_Q (l_Q+2 l) (85 \mathcal{E}^2+80 \mathcal{E}+8) \Big] \,. \nonumber
\end{align}
In the equatorial limit $l_Q \to l${,} we find perfect agreement with the expression in the literature \cite{Kalin:2019inp,Kalin:2019rwq}.

\subsection{The orbital plane precession rate $\omega_{\theta}/\omega_r$}

{The orbital plane precession is determined by $K^{\phi r}-K^{\theta r}$, whose leading weak-field term gives the Lense-Thirring effect.}  Using \eqref{eq:ratiotheta}, we obtain up to $\mathcal{O}(M^3 a^2)$
\begin{align}
\label{eq:Kthetar}
&\hspace{-10pt}K^{\theta r} =1 + \frac{3 M^2 (5 \mathcal{E}+4)}{4 \, l_Q^2}
-\frac{3 a M^2 l \sqrt{\mathcal{E}+1} (5 \mathcal{E}+2)}{l_Q^4} \\
& \qquad + \frac{3 a^2 M^2}{32 \, l_Q^6} \Big[ l^2(445 \mathcal{E}^2+416 \mathcal{E}+40) \nonumber \\
& \qquad \qquad \qquad \qquad  \qquad -  l_Q^2 (85 \mathcal{E}^2+80 \mathcal{E}+8) \Big] \,. \nonumber
\end{align}
In the limit $a \to 0${,} the azimuthal and polar frequencies in \eqref{eq:Kphir}--\eqref{eq:Kthetar} become degenerate $\omega_{\theta} = \omega_{\phi}$, {while exactly equatorial orbits have no polar oscillation.}

\section{Conclusion}

In this paper, we have explored the relationship between scattering and bound observables for generic orbits in a Kerr background. The establishment of a boundary-to-bound dictionary represents a crucial step towards leveraging the computational tools that have been developed for scattering amplitudes in the study of bound systems. Expanding upon previous work in the field, we have extended such dictionary beyond the equatorial case by considering a smooth class of geodesics which interpolate between scattering and bound dynamics.

Taking advantage of the Hamilton-Jacobi representation, we have been able to write down a closed form expression for the radial and the polar contribution to the action for such generic class of scattering and bound orbits. In particular, we have found that in the PM expansion there is one turning point in the scattering case and two turning points in the bound case both for the radial and the polar motion. Such analytic continuation involves also the Carter constant, which plays a crucial role for the dynamics beyond the equatorial plane.

We have then computed, in the post-Minkowskian expansion, the azimuthal $\Delta \phi$ and the polar $\Delta \theta$ deflection angles for time-like geodesics in Kerr and the time delay $\Delta T$ for null geodesics. While the azimuthal angle is naturally derived from the action, the polar angle has a more implicit expression since there is no natural conjugate variable. Indeed{,} in the conventional partial-wave basis{,} an explicit relation has been found only for some degenerate configurations \cite{Glampedakis:2001cx,Bautista:2021wfy}, but perhaps an alternative basis might help to clarify the general case \cite{Kol:2021jjc}. Using the new boundary to bound dictionary, we have then studied the weak-field expansion of the periastron and the orbital plane precession rate, $K^{\phi r}$ and ${K^{\phi r}-K^{\theta r}}$, which are uniquely defined from the ratio of fundamental frequencies \cite{Schmidt:2002qk}.

This work offers new promising directions for the analytic continuation of classical scattering and bound observables beyond the equatorial case. First, it would be important to extend the amplitude-action relation for generic angular momentum orientations, which would include some type of polar action contribution. Furthermore, a natural extension of our work would be to consider a spinning probe in a Kerr background \cite{Witzany:2019nml}, since a generalization of the Carter constant has been discovered by R\"udiger in the pole-dipole approximation \cite{rudiger1981conserved, rudiger1983conserved} and recently generalized to quadrupolar order~\cite{Compere:2023alp}. Finally, it would be interesting to see how the extension of the Schwinger-Dyson recursion \cite{Adamo:2022ooq}  would allow to compute radiative observables for bound orbits. We hope to come back to these questions in the near future.

\vspace{-10pt}
\begin{acknowledgments}
We are very grateful to G. K\"alin, D. Kosmopoulos, A. Ilderton and J. Vines and R. Porto for discussions and useful comments on the draft. CS is funded by China Postdoctoral Science Foundation under Grant NO. 2022TQ0346.
\end{acknowledgments}

\appendix

\section{Conserved charges and kinematics}
\label{sec:AppendixD}
Here we present the relations between the conserved quantities and the kinematic invariants in the scattering case. Consider a probe particle with incoming momentum $p^{\mu}_{\text{in}} = (E, \vec{p}_\text{in})$ and impact parameter $b^\mu = (0, \vec{b})$, defined in such a way that $\vec{p}_\text{in} \cdot \vec{b} = 0$. Then we have
\begin{gather}
    E^2 = |\vec{p}_{\text{in}}|^2 + m^2, \nonumber \\
    L^2 = |\vec{p}_{\text{in}}|^2 \big(|\vec{b}|^2 - (\hat{a} \cdot \vec{b})^2 \big) - |\vec{b}|^2 (\hat{a} \cdot \vec{p}_{\text{in}})^2 , \nonumber \\
    Q = (|\vec{b}|^2 - |\vec{a}|^2)(\hat{a} \cdot \vec{p}_{\text{in}})^2 + |\vec{p}_{\text{in}}|^2 (\hat{a} \cdot \vec{b})^2 ,
    \label{eq:kinematics}
\end{gather}
where $\hat{a} = {\vec{a}}/{|\vec{a}|}$ is the unit vector along the spin direction. The incoming $\theta$ angle is determined by
\begin{align}
    U_\mathrm{in} = \cos^2(\theta_\mathrm{in}) = (\hat{a} \cdot \hat{p}_{\text{in}})^2 \,.
\end{align}
In the case considered in this letter{,} $\theta_\mathrm{in} = \pi/2$ and therefore we have imposed $\hat{a} \cdot \hat{p}_{\text{in}}=0$.

\section{Turning points of the polar motion}
\label{sec:AppendixA}

The $(r,\theta)$ components of the geodesic equations in a Kerr black hole imply
\begin{align}
\int_{\mathcal{C}_r} \frac{\mathrm{d}r}{\sqrt{R(r)}} = \int_{\mathcal{C}_{\theta}} \frac{\mathrm{d}\theta}{\sqrt{\Theta(\theta)}} \,,
\label{eq:rtheta_eq}
\end{align}
from which we can find the final polar angle $U_{\text{out}}$, as discussed in section IID of \cite{Kapec:2019hro}. A direct calculation for our setup shows that \eqref{eq:rtheta_eq} can be reduced to
\begin{align}
\label{eq:turning_points_scattering}
&\sqrt{-a^2 m^2 \mathcal{E} U_-} \int_{\mathcal{C}_r^{>}} \frac{\mathrm{d}r}{\sqrt{R(r)}}  \\
&= 2 n^{>} K \left(\frac{U_+}{U_-}\right) {+} F\left(\arcsin\left(\sqrt{\frac{U_{\text{out}}^{>}}{U_+}}\right)\Big|\frac{U_+}{U_-}\right) \nonumber
\end{align}
for the scattering case $\mathcal{E} > 0$ and to
\begin{align}
\label{eq:turning_points_bound}
&\sqrt{-a^2 m^2 \mathcal{E} U_-} \int_{\mathcal{C}_r^{<}} \frac{\mathrm{d}r}{\sqrt{R(r)}}   \\
&= 2 n^{<} K \left(\frac{U_+}{U_-}\right) + F\left(\arcsin\left(\sqrt{\frac{U_{\text{out}}^{<}}{U_+}}\right)\Big|\frac{U_+}{U_-}\right) \nonumber
\end{align}
in the bound case $\mathcal{E} < 0$. {The plus signs reflect the fact that $\eta_{\text{out}} = -1$ for both scattering and bound orbits on the branches considered here.} Since $U_{\text{out}}$ is independent of $n$, it turns that a perturbative expansion of \eqref{eq:turning_points_scattering} and \eqref{eq:turning_points_bound} completely fix the number of turning points in the polar motion {on these branches} to
\begin{align}
n^{>} = 1\,, \qquad n^{<} = 2 \,.
\end{align}

\begin{figure}[h!]
\begingroup
\setlength{\unitlength}{\linewidth}
\begin{picture}(1,0.590724638)
\put(0,0){\includegraphics[width=\linewidth]{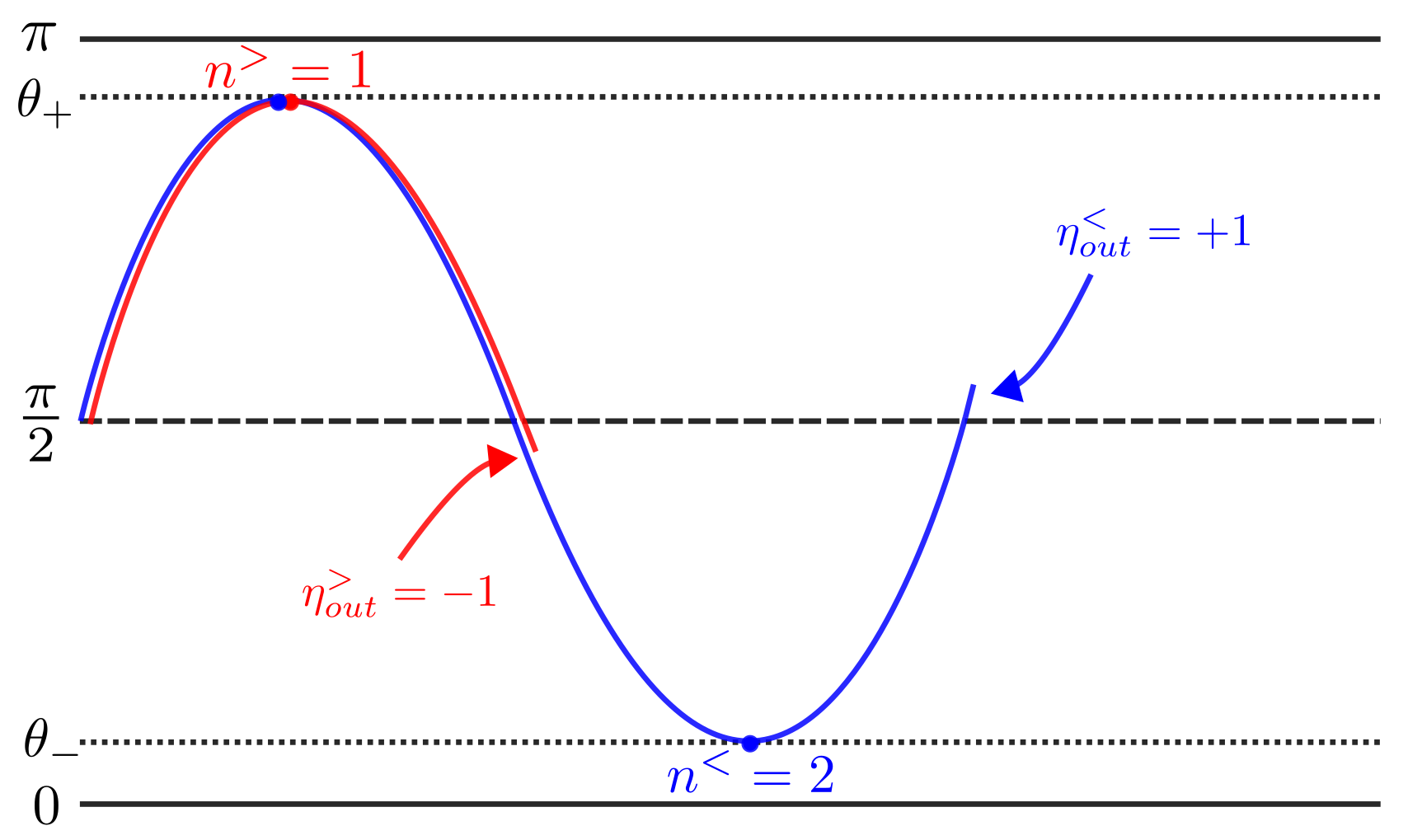}}
\put(0.840000,0.413913){\color{white}\rule{0.026087\linewidth}{0.025507\linewidth}}
\put(0.841739,0.425758143){\color{blue}\rule{0.022029\linewidth}{0.001236715\linewidth}}
\end{picture}
\endgroup
\caption{A pictorial representation of the polar motion for the scattering (in red) and bound (in blue) class of geodesics of interest in this paper.}
\label{fig:polar_angle}
\end{figure}

\section{Exact expressions for scattering observables}
\label{sec:AppendixB}

We provide here some compact resummed expressions for scattering observables in terms of hypergeometric functions (see also \cite{Kraniotis:2005zm} for the equatorial case). These are always functions of the roots of the radial and polar potentials, which needs to be explicitly derived for the perturbative calculations.

The polar deflection angle is given by \eqref{eq:Uout}
\begin{align}
&\qquad\cos^2(\theta_{\text{out}}) = U_{+} \operatorname{sn}^2\left(X_0^{>} \Big| {\frac{U_{+}}{U_{-}}}\right) \,,\qquad \mathcal{E}>0\,, Q>0\,, \nonumber \\
&X_0^{>}=- 4 u_m \sqrt{-a^2 U_{-}} \, F_D^{(3)}\left(\alpha_{\Delta\theta},\vec{\beta}_{\Delta\theta},\gamma_{\Delta\theta}; \frac{u_m}{u_1},\frac{u_m}{u_2},\frac{u_m}{u_3}\right)  \,, \nonumber \\
&\qquad\,\,\alpha_{\Delta\theta}=1, \quad \vec{\beta}_{\Delta\theta} = \Big\{\frac{1}{2},\frac{1}{2},\frac{1}{2}\Big\}, \quad \gamma_{\Delta\theta}=\frac{3}{2}\,,
\end{align}
\par\noindent\begin{minipage}{\linewidth}
while the azimuthal deflection angle is derived from \eqref{eq:phi_angle}
\begingroup\small
\begin{align}
&\Delta\phi =\! {2} u_m \Bigg[  \frac{G_A}{u_A} F_D^{(4)}\left(\alpha_{\Delta\phi_1},\vec{\beta}_{\Delta\phi_1},\gamma_{\Delta\phi_1}; \frac{u_m}{u_A},\frac{u_m}{u_1},\frac{u_m}{u_2},\frac{u_m}{u_3}\right) \nonumber \\
&-\frac{G_B}{u_B} F_D^{(4)}\left(\alpha_{\Delta\phi_1},\vec{\beta}_{\Delta\phi_1},\gamma_{\Delta\phi_1}; \frac{u_m}{u_B},\frac{u_m}{u_1},\frac{u_m}{u_2},\frac{u_m}{u_3}\right) \Bigg]\nonumber \\
& + \frac{\pi\, l\, {U_+}^{\frac{3}{2}}}{2 \sqrt{l_Q^2 -l^2}} F_D^{(2)}\left(\alpha_{\Delta\phi_2},\vec{\beta}_{\Delta\phi_2},\gamma_{\Delta\phi_2}; {U_+},\frac{{U_+}}{{U_-}}\right) \nonumber \\
& + l \frac{(U^{>}_{\text{out}})^{\frac{3}{2}}}{3 \sqrt{l_Q^2 -l^2 }} F_D^{(3)}\left(\alpha_{\Delta\phi_3},\vec{\beta}_{\Delta\phi_3},\gamma_{\Delta\phi_3}; U^{>}_{\text{out}},\frac{U^{>}_{\text{out}}}{{U_-}},\frac{U^{>}_{\text{out}}}{{U_+}}\right) {-\pi} \,,\nonumber \\
&G_A = \frac{2 M u_A (l -a {\sqrt{1+\mathcal{E}}}) - l}{\sqrt{\mathcal{E} (M^2 - a^2)}}\,,\nonumber \\
&G_B = \frac{2 M u_B (l - a {\sqrt{1+\mathcal{E}}}) - l}{\sqrt{\mathcal{E} (M^2 - a^2)}}\,,\nonumber \\
&\qquad  \alpha_{\Delta\phi_1}=1, \quad \vec{\beta}_{\Delta\phi_1} = \Big\{1,\frac{1}{2},\frac{1}{2},\frac{1}{2}\Big\}, \quad \gamma_{\Delta\phi_1}=\frac{3}{2}\,, \nonumber \\
&\qquad  \alpha_{\Delta\phi_2}=\frac{3}{2}, \quad \vec{\beta}_{\Delta\phi_2} = \Big\{1,\frac{1}{2}\Big\}, \quad \gamma_{\Delta\phi_2}=2\,, \nonumber \\
&\qquad \alpha_{\Delta\phi_3}=\frac{3}{2}, \quad \vec{\beta}_{\Delta\phi_3} = \Big\{1,\frac{1}{2},\frac{1}{2}\Big\}, \quad \gamma_{\Delta\phi_3}=\frac{5}{2}\,.
\end{align}
\endgroup
\end{minipage}\par

\section{Derivation of the fundamental frequencies}
\label{sec:AppendixC}

The four integrals of motion
\begin{align}
P_{\alpha}=\left(H, E, L, Q\right)=\left(-\frac{1}{2} m^2, E, L, Q\right) \,
\end{align}
are implicit functions of the action variables $P_{\alpha} = f(J_{\beta})$
\begin{align}
J_{\beta} = (J_t, J_r, J_{\theta}, J_{\phi}) \,.
\end{align}
The fundamental frequencies can therefore be computed from the jacobian of $f$, i.e.{,}
\begin{align}
\frac{\partial P_{\alpha}}{\partial J_{\beta}} \frac{\partial J_{\beta}}{\partial P_{\gamma}} = \delta_{\alpha}^{\gamma} \,,
\end{align}
which gives directly \eqref{eq:fund_frequencies}.
\\
~
\\
~
\\

\nocite{Damour1985}
\bibliography{apssamp}

\end{document}